\documentclass[letter]{aa} 
\usepackage{graphicx} 
\usepackage{times}
\usepackage{natbib} 
\usepackage{txfonts}
\bibpunct{(}{)}{;}{a}{}{,} 

\newcommand{\mearth}{$\mathrm{{M}_\oplus}$}
\newcommand{\mearths}{$\mathrm{{M}_\oplus}$ }

\begin{document}

\title{Halting Type I planet migration in non-isothermal disks} 
\author{Sijme-Jan Paardekooper \inst{1} \and Garrelt Mellema \inst{2,1}} 
\offprints{S. J. Paardekooper} 
\institute{Leiden Observatory, Postbus 9513, NL-2300 RA Leiden, 
           The Netherlands \\
	   \email{paardeko@strw.leidenuniv.nl} \and
           Stockholm Observatory, AlbaNova University Center,
	   Stockholm University, SE-106 91 Stockholm, Sweden \\
	   \email{garrelt@astro.su.se}} 

\date{Draft Version \today} 
  
\abstract{}{We investigate the effect of including a proper energy
  balance on the interaction of a low-mass planet with a
  protoplanetary disk.}{We use a three-dimensional version of the
  RODEO method to perform hydrodynamical simulations including the
  energy equation. Radiation is included in the flux-limited diffusion
  approach.}{The sign of the torque is sensitive to the ability
  of the disk to radiate away the energy generated in the immediate
  surroundings of the planet. In the case of high opacity,
  corresponding to the dense inner regions of protoplanetary disks,
  migration is directed \emph{outward}, instead of the usual inward
  migration that was found in locally isothermal disks. For low values
  of the opacity we recover inward migration and show that torques
  originating in the coorbital region are responsible for the change
  in migration direction.}{} 
    
  
\titlerunning{Halting Type I planet migration}

\maketitle

  
\section{Introduction}
Migration has become a standard ingredient in planet formation
theory, providing an explanation for the extrasolar giant planets that
were found orbiting very close to their central star, the so-called
Hot Jupiters. It is believed that these planets were formed at a
distance of several Astronomical Units (AU), after which they moved in
due to tidal interaction with the surrounding protoplanetary disk.

Depending on the mass of the planet, three different migration modes
can be distinguished:
\begin{enumerate}
\item Type I migration
  \citep{1979ApJ...233..857G,1997Icar..126..261W}, valid for low-mass 
  planets. The disk response is linear in this case, and the resulting
  torques can be calculated in a semi-analytic way
  \citep{2002ApJ...565.1257T}. Planets that are less massive than a
  few Earth masses are subject to this mode of migration. 
\item Type II migration, for which the planet orbits in a deep annular
  gap \citep{1986ApJ...309..846L}. In this case, the planet is tidally
  locked inside the gap, and the time scale for inward migration
  equals the viscous time scale, which approximately equals $10^6$
  yr. This mode of migration holds for planets comparable to or more 
  massive than Jupiter. See, however, \cite{2002ApJ...572..566R} on
  gap formation for   low-mass planets.
\item In between these two regimes, interesting things happen. The
  onset of non-linear behavior can influence the torque to a large
  extent \citep{2006astro.ph..7155M}. However, when the disk is
  massive enough, the planet enters a new migration mode that can be
  very fast and that relies on dynamical corotation torques
  \citep{2003ApJ...588..494M,2004ASPC..324...39A}. This is called Type
  III migration, and it holds for planets that open up a partial gap.
\end{enumerate}
In this Letter, we focus on deeply embedded low-mass planets and,
therefore, on the regime of Type I migration. 

The time scale for Type I migration is inversely proportional to the
mass of the disk and the planet \citep{2002ApJ...565.1257T}, and it can
be much shorter than the disk lifetime of approximately $10^7$ yr. A
planet of 1 Earth mass (1 \mearth), for example, embedded in the
minimum mass solar nebula (MMSN) at 5 AU, would migrate into the
central star within $10^6$ yr. Within the widely accepted
core-accretion model of giant planet formation
\citep{1996Icar..124...62P}, the gas giants also pass through a stage
where the core is a few \mearth, which thus should disappear into the
central star even faster. Therefore, some sort of stopping mechanism
is required for planets to survive Type I migration.

Proposed stopping mechanisms include magnetic turbulence
\citep{2004MNRAS.350..849N}, a toroidal magnetic field
\citep{2003MNRAS.341.1157T}, and an inner hole in the disk
\citep{2002ApJ...574L..87K,2006ApJ...642..478M}. However,
most analytical and numerical work on planet migration has 
focused on disks with a fixed radial temperature profile. For these
disks, hydrodynamical simulations can reproduce Type I migration in
both two and three dimensions \citep{2002A&A...385..647D,
2003ApJ...586..540D, 2003MNRAS.341..213B}. 

The isothermal assumption is only valid if the disk can radiate away
all excess energy efficiently. When the disk is optically thick, the
radiative energy flux through the surface of a sphere of radius $H$
around the planet is given by $F_\mathrm{R}=\sigma T^4/\tau$, where
$\sigma$ is the Stefan-Boltzmann constant and $\tau$ is the optical
depth over a distance $H$. Therefore, when the opacity increases, less
energy can be radiated away. This makes the cooling time scale much
longer than the dynamical time scale in the dense inner regions of
protoplanetary disks, and therefore the isothermal assumption is
likely to be invalid there.

One may wonder what effects releasing the isothermal assumption may
have on the torque. The effects of a more realistic temperature
profile on the Lindblad torques were investigated by
\cite{2005ApJ...619.1123J} and \cite{2004ApJ...606..520M}, who found
that the migration rate is very sensitive to the temperature and
opacity structure and that in some cases migration may be very slow
compared to the isothermal Type I case. In this Letter, we present the
first results of including a detailed energy balance on the migration
behavior of low-mass planets in a radiation-hydrodynamical context. In 
Sect. \ref{secMod} we discuss the disk model, in Sect. \ref{secRes} we
present the results, and we give a discussion of the results in Sect. 
\ref{secDis}. We then conclude in Sect. \ref{secCon}. 
 
  
\section{Disk model}
\label{secMod}
We work in spherical coordinates $(r,\theta,\phi)$ with the
central star in the origin, and the unit of distance is the planet's
orbital radius. In the plots we also use a Cartesian coordinate
frame with the central star in the origin and the planet located at
$(x,y,z)=(-1,0,0)$. The planet stays on a fixed circular orbit
throughout the simulation. The disk is three-dimensional, and the
computational domain is bounded by $r=0.4$, $r=2.5$, $\theta=\pi/2$,
and $\theta=\pi/2-5h/2$, where $h$ is the relative scale height of the 
disk ($h=H/r$, with $H$ the pressure scale height). The disk spans the
full $2\pi$ in azimuth. For the radial and meridional boundaries, we use
non-reflective boundary conditions \citep{2006A&A...450.1203P}, except
for the boundary at $\theta=\pi/2$, which is fully reflective. The
azimuthal boundary conditions are periodic. This domain is covered by
a grid with 256 radial cells, 768 azimuthal cells, and 16 meridional
cells. This configuration leads to cubic cells near the planet. On top
of this main grid, we put 5 levels of adaptive mesh refinement (AMR)
near the planet. Each level increases the local resolution by a factor
of 2, which leads to a local resolution of $\Delta r \approx
0.00025$. We focus on a planet of 5 \mearths located at 5 AU from the
central star, for which the Hill radius is then covered by 70
cells. While this resolution is not needed for accurate torque 
calculations, it is neede for correctly modeling the planet's envelope
and for the process of accretion, which we will consider in a future
publication. In the torque calculations, we neglect the material orbiting
inside the Hill sphere. For deeply embedded objects, material
could be bound in the smaller Bondi sphere
\citep{2006ApJ...642..478M}, so that we may exclude a region
that is too large. However, we checked that the results are not
sensitive to this.

The disk is taken to be inviscid, so as to filter out effects of viscous
heating. The density follows a power law initially with index
$-3/2$. We take $h=0.05$, a typical value used in simulations of
disk-planet interaction. This sets the temperature at approximately 50
K at 5 AU. The disk is in Keplerian rotation initially with a slight
correction for the radial pressure gradient. The potential contains
terms due to the star, the planet, and the acceleration of the
coordinate frame due to the planet and the disk. The last two
contributions are of negligible importance for a planet of 5
\mearth. We smooth the potential of the planet over 2 grid cells. 

We solve the flow equations using a three-dimensional version of the
RODEO method \citep{2006A&A...450.1203P}, with an energy equation
included. Radiative transfer is treated in the flux-limited diffusion
approach \citep{1981ApJ...248..321L}, together with the flux limiter
of \cite{1989A&A...208...98K}. The three-dimensional hydrodynamics
solver (including AMR) was tested through isothermal disk-planet
interaction, which showed that we could reproduce Type I and Type II
migration in the relevant mass regime. The radiative transfer module
was tested against multi-dimensional diffusion problems. We used
opacity data from \cite{1994ApJ...427..987B}. In the temperature range
of interest, the opacity is proportional to the density and
temperature squared. We vary the initial midplane density to study
different cooling regimes while keeping the initial temperature fixed.  

The code is parallelized using MPI, and the simulations were run on
126 $1.3$ GHz processors at the Dutch National Supercomputer. Each
radiation-hydrodynamical model took approximately 4 days of computing
time. 

  
\section{Results}
\label{secRes}
\begin{figure}
\centering
\resizebox{\hsize}{!}{\includegraphics[bb=248 9 501 245,clip]{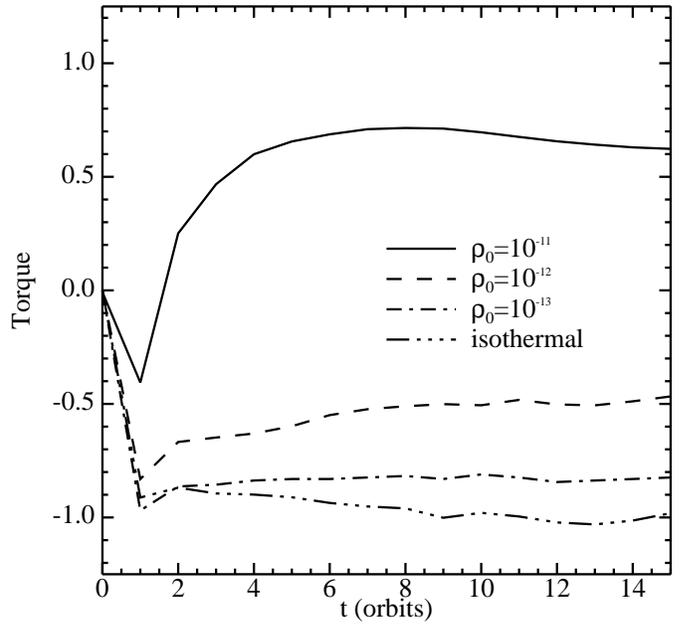}}
\caption{Total torque on a 5 \mearths planet as a function of time for
  three different midplane densities, together with the isothermal
  result. The torques are normalized to the analytical value found by
  \cite{2002ApJ...565.1257T}, which is reproduced by the isothermal
  simulation. For high densities (and thereby for high opacities) the
  torque becomes positive, indicating outward migration.}
\label{fig1}
\end{figure}

In Fig. \ref{fig1} we show the evolution of the total torque on our 5
\mearths planet for three different midplane densities, where
$\rho_0=10^{-11}$ $\mathrm{g~cm^{-3}}$ corresponds to the MMSN at 5
AU. For this high density, the torque is \emph{positive}, indicating
outward migration. The absolute value of the torque is approximately a 
factor 2 lower than the Type I torque. For lower densities, and
therefore for lower opacities, we recover inward migration; and for
the lowest value of the density, we approach inward Type I migration of
isothermal disks.  

\begin{figure}
\centering
\resizebox{\hsize}{!}{\includegraphics[bb=248 9 505 245,clip]{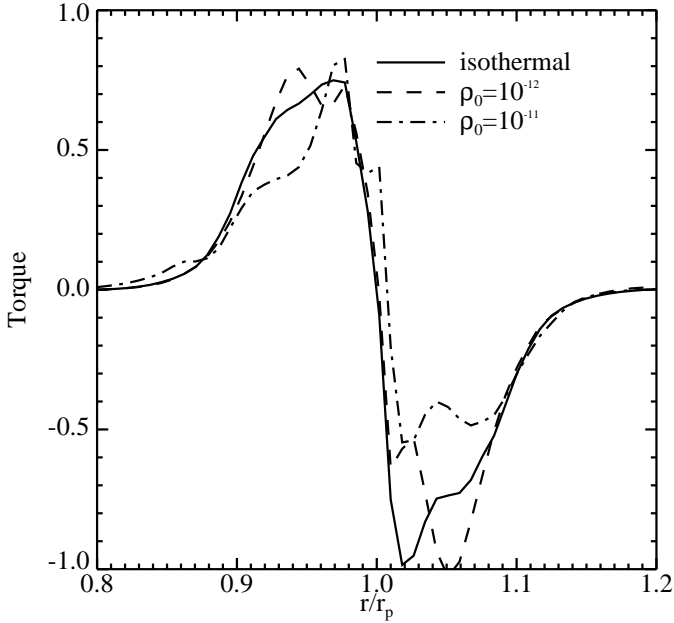}}
\caption{Radial torque distribution for a 5 \mearths planet for two
  different densities, together with the result for a locally
  isothermal equation of state. Both the inner wave and the outer wave
  are reduced in strength for higher densities, but the major
  difference comes from the corotation region.}
\label{fig2}
\end{figure}

To find out where this positive torque originates, we show the radial
torque profile close to the planet in Fig. \ref{fig2}. An embedded
planet excites two spiral waves into the disk. The inner spiral wave
is responsible for the positive bump near $r=0.95$, while the outer
spiral wave creates the negative bump near $r=1.05$. The location and
the strength of these bumps varies between the different runs, but the
total Lindblad torque remains negative for all simulations. For
$\rho_0=10^{-11}$ $\mathrm{g~cm^{-3}}$, the magnitude of the Lindblad
torque is a factor of 2 smaller compared to the isothermal case. This
implies that it is the corotation region, located between $r=0.95$ and
$r=1.05$, that is responsible for the change in sign of the total
torque.   

\begin{figure}
\centering
\resizebox{\hsize}{!}{\includegraphics[bb=248 9 501 245,clip]{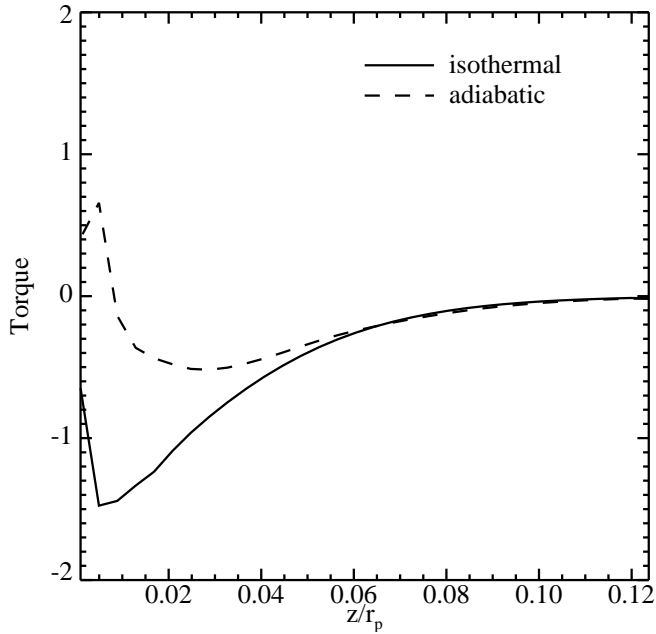}}
\caption{Vertical torque distribution for an adiabatic simulation with
no radiative cooling compared to the locally isothermal result. The
difference starts high above the Hill sphere of the planet, which is
approximately at $z=0.017$.}
\label{fig3}
\end{figure}

\begin{figure}
\centering
\resizebox{\hsize}{!}{\includegraphics[]{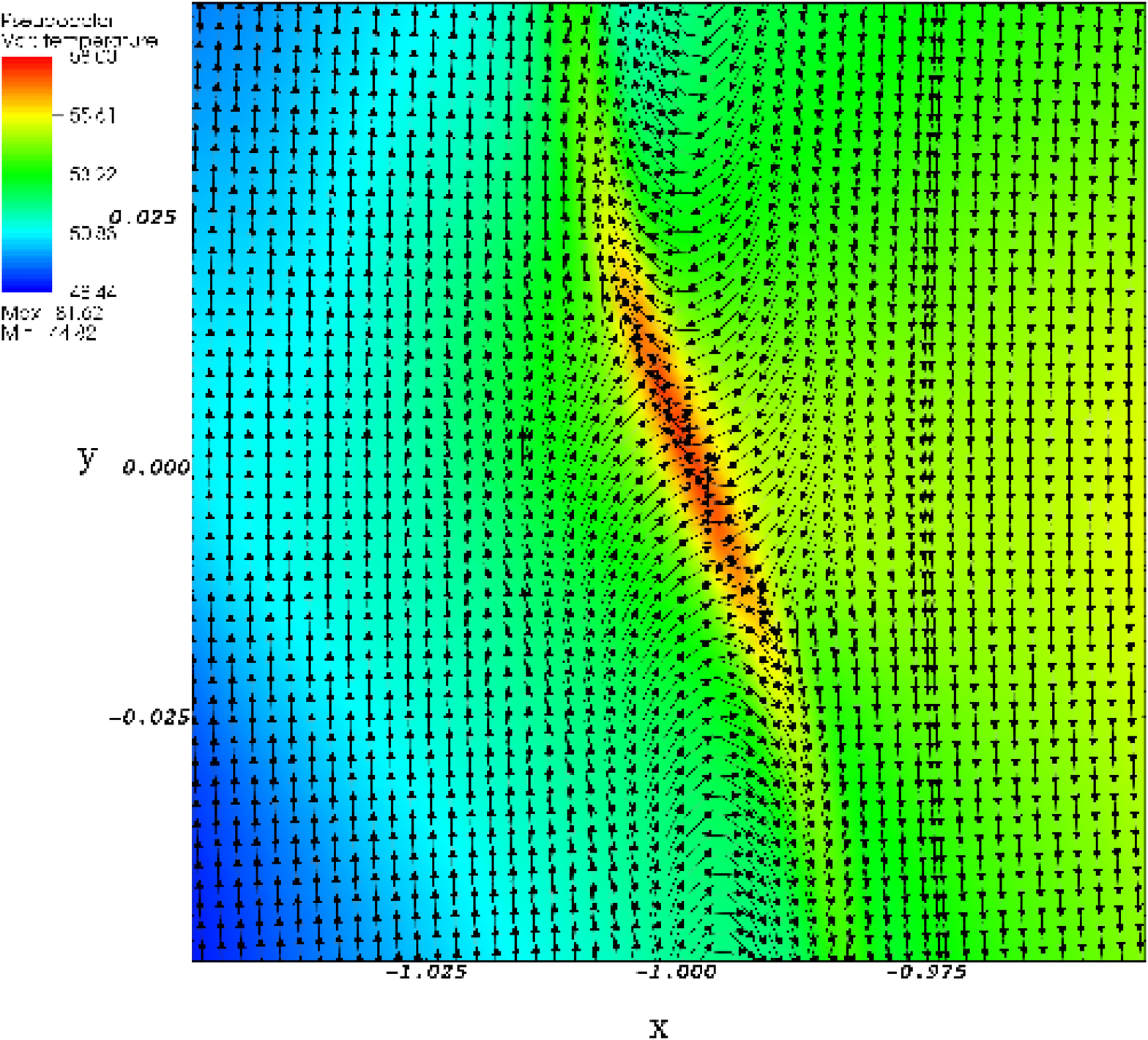}}
\caption{Temperature at $z=0.02$, slightly above the Hill sphere of
  the planet. Overplotted is the direction of the velocity field,
  clearly showing the horseshoe orbits. The point where the two
  horseshoe legs meet is a region of compression, meaning the
  temperature is slightly higher at that location.}
\label{fig4}
\end{figure}

We can pin down the origin of the positive corotation torque further 
by looking at the vertical torque distribution. In Fig. \ref{fig3} we
compare an isothermal run to an adiabatic simulation. The latter does
not include radiative cooling, so the effect of inefficient
cooling is even more pronounced. The important thing to note is that
the difference in Fig. \ref{fig3} starts well above the Hill sphere of
the planet, the edge of which is located at approximately
$z=0.017$; therefore, we can filter out the effects of heating deep
within the planetary envelope and look at temperature differences
above $z=0.017$. 

\begin{figure}
\centering
\resizebox{\hsize}{!}{\includegraphics[]{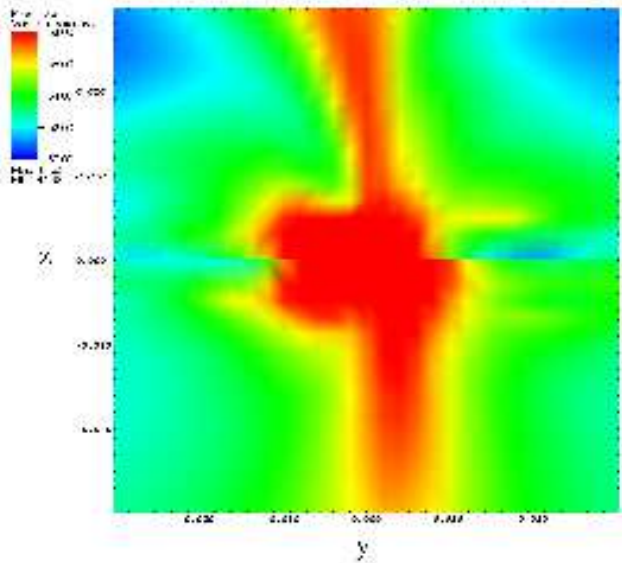}}
\caption{Temperature at $x=-1$ (or $r=1$), showing the hot plume
  behind the planet ($y>0$, the planet moves towards the left of the
  figure). This plume extends all the way to $z=0$, but there the  
  heating inside the planetary envelope dominates the temperature
  structure. For $z>0$ we show a simulation without a global
  pressure gradient (i. e. the gas orbits at the Kepler velocity),
  while for $z<0$ we show a simulation with a global radial pressure
  gradient ($p\propto \rho T \propto r^{-2}$). The difference in the
  position of the warm plume is apparent for both simulations. The
  vertical line at $y=0$ is there to guide the eye. The color scale
  has been chosen to focus on the warm plumes.} 
\label{fig5}
\end{figure}

We show a cut through $z=0.02$ in Fig. \ref{fig4}, showing the
temperature and the direction of the velocities. The horseshoe orbits
clearly stand out, and where the two horseshoe legs meet, the gas is
compressed and the temperature rises. Due to the slightly
sub-Keplerian velocity of the gas, the two horseshoe legs meet 
\emph{behind} the planet (see Fig. \ref{fig5}). As the disk tries to
maintain pressure equilibrium, the density will be lower in this
region. We verified that the temperature rise is enough to
account for the positive torque by verifying that 
\begin{equation}
\int_\mathrm{disk} \frac{\mathrm{G}M_\mathrm{p}\partial \rho}{|{\vec
    r} - {\vec r_\mathrm{p}}|^3} {\vec{r_\mathrm{p}}} \times ({\vec
    r}-{\vec r_\mathrm{p}}) {\vec d^3r},
\end{equation}
with $\partial \rho=-\rho_0 \partial T/T_0$, is within the same order of
magnitude as the total torque on the planet. Equation (1) reproduced
the total torque to within 25 \%, which shows that indeed the
temperature change is strong enough to induce a density asymmetry that
is capable of producing the observed positive torque. It is this
density asymmetry that leads to a large, \emph{positive} contribution
to the torque. Only if the disk can radiate away this heat efficiently
does the total torque become negative again.  


\section{Discussion}
\label{secDis}
The corotation region is very sensitive to local conditions
\citep{2001ApJ...558..453M,2002A&A...387..605M}, including the shape
of the planetary envelope. This may affect the contribution to the
total torque. However, we have found that the positive corotation
torque is a robust effect, showing no strong dependence on numerical
resolution, for example. Because we take a smoothing length of 2 grid
cells, we have also found no dependence on the exact form of the
planetary potential. 

We estimate that the transition between inward and outward migration
happens around 15 AU in the MMSN, based on the simulations with
different opacities. Although we parametrized the opacity using
the local gas density, it is the opacity that matters. If the opacity
is lowered, as when it is due to grain growth
\citep{2005A&A...434..971D}, the transition radius will be located
farther inward.   

The computational costs of the radiation-hydrodynamical simulations is
too high to run simulations for hundreds of orbits. Therefore, effects
that happen on the libration time scale, such as saturation of the
corotation torque \citep{2003ApJ...587..398O}, are not captured by
these simulations. Adiabatic runs showed that indeed saturation occurs
on the libration time scale, which considerably reduces the magnitude of the
positive torque. This is another confirmation that is
indeed the corotation region that is responsible for the positive
torque. This does not mean, however, that the positive torque should
disappear on a libration time scale. When the radiation diffusion time
scale is shorter than the libration time scale, the asymmetry in
temperature, and therefore in density, can be maintained. In the MMSN,
this is the case for approximately $r>1$ AU. 

For lower-mass planets, the horseshoe region shrinks, so
the positive torque should decrease in strength. However, in this case
the Lindblad torques also decrease in magnitude. We have verified that
even for a planet of $0.5$ \mearths the total torque is still positive
for $\rho_0=10^{-11}$ $\mathrm{g~cm^{-3}}$. Further study is required
to pin down the exact dependence of the torque on planetary mass. We 
expect that in the high-mass regime, when the planet starts to open up
a gap and when the density around the planet decreases by several orders
of magnitude, cooling will always be efficient. Therefore, Type II
migration should not be very different from the isothermal case
\citep[see also][]{2006A&A...445..747K}. 

  
\section{Conclusions}
\label{secCon}
In this Letter, we have presented the first radiation-hydrodynamical
simulations of low-mass planets embedded in a protoplanetary disk. We
find that the migration behavior of a 5 \mearths planet depends
critically on the ability of the disk to radiate away the heat
generated by compression close to the planet. When radiative cooling
is not efficient, the torque on the planet is \emph{positive},
indicating outward migration. Analysis of the torque distribution
showed that it is the coorbital region that generates the large
positive contribution to the torque, while the Lindblad torque is
reduced in strength, but is still negative. Local, adiabatic simulations
showed that compression at the turn-over point of the horseshoe orbits
creates a warm region behind the planet. As the disk tries to maintain
pressure equilibrium, the density behind the planet will become lower,
which leads to a positive corotation torque. This happens in the dense
inner parts of protoplanetary disks, where the opacity is high. In
regions of low opacity, we recover inward Type I migration. Therefore,
low-mass planets do not migrate inward all the way to the central
star, bun instead stop migrating when radiative cooling becomes
inefficient. In the MMSN, the transition occurs at approximately 15
AU. As the disk slowly accretes onto the central star, this radius
will move inward; as a result, low-mass planets will also 
eventually continue their inward migration, but on the much longer
viscous time scale. This way, low-mass planets are saved from fast
inward Type I migration into the central star.

  
\begin{acknowledgements}
We thank the anonymous referee for an insightful report. SP thanks
Yuri Levin, Mordecai-Marc MacLow, and Peter Woitke for useful
discussions. We thank Willem Vermin for his assistance at the Dutch
National Supercomputer. This work was sponsored by the National
Computing Foundation (NCF) for the use of supercomputer facilities,
with financial support from the Netherlands Organization for
Scientific Research (NWO). 
\end{acknowledgements}

  
\bibliographystyle{aa} 
\bibliography{planet1.bib}

\begin{thebibliography}{27}
\expandafter\ifx\csname natexlab\endcsname\relax\def\natexlab#1{#1}\fi

\bibitem[{{Artymowicz}(2004)}]{2004ASPC..324...39A}
{Artymowicz}, P. 2004, in ASP Conf. Ser. 324: Debris Disks and the Formation of
  Planets, 39

\bibitem[{{Bate} {et~al.}(2003){Bate}, {Lubow}, {Ogilvie}, \&
  {Miller}}]{2003MNRAS.341..213B}
{Bate}, M.~R., {Lubow}, S.~H., {Ogilvie}, G.~I., \& {Miller}, K.~A. 2003,
  \mnras, 341, 213

\bibitem[{{Bell} \& {Lin}(1994)}]{1994ApJ...427..987B}
{Bell}, K.~R. \& {Lin}, D.~N.~C. 1994, \apj, 427, 987

\bibitem[{{D'Angelo} {et~al.}(2002){D'Angelo}, {Henning}, \&
  {Kley}}]{2002A&A...385..647D}
{D'Angelo}, G., {Henning}, T., \& {Kley}, W. 2002, \aap, 385, 647

\bibitem[{{D'Angelo} {et~al.}(2003){D'Angelo}, {Kley}, \&
  {Henning}}]{2003ApJ...586..540D}
{D'Angelo}, G., {Kley}, W., \& {Henning}, T. 2003, \apj, 586, 540

\bibitem[{{Dullemond} \& {Dominik}(2005)}]{2005A&A...434..971D}
{Dullemond}, C.~P. \& {Dominik}, C. 2005, \aap, 434, 971

\bibitem[{{Goldreich} \& {Tremaine}(1979)}]{1979ApJ...233..857G}
{Goldreich}, P. \& {Tremaine}, S. 1979, \apj, 233, 857

\bibitem[{{Jang-Condell} \& {Sasselov}(2005)}]{2005ApJ...619.1123J}
{Jang-Condell}, H. \& {Sasselov}, D.~D. 2005, \apj, 619, 1123

\bibitem[{{Klahr} \& {Kley}(2006)}]{2006A&A...445..747K}
{Klahr}, H. \& {Kley}, W. 2006, \aap, 445, 747

\bibitem[{{Kley}(1989)}]{1989A&A...208...98K}
{Kley}, W. 1989, \aap, 208, 98

\bibitem[{{Kuchner} \& {Lecar}(2002)}]{2002ApJ...574L..87K}
{Kuchner}, M.~J. \& {Lecar}, M. 2002, \apjl, 574, L87

\bibitem[{{Levermore} \& {Pomraning}(1981)}]{1981ApJ...248..321L}
{Levermore}, C.~D. \& {Pomraning}, G.~C. 1981, \apj, 248, 321

\bibitem[{{Lin} \& {Papaloizou}(1986)}]{1986ApJ...309..846L}
{Lin}, D.~N.~C. \& {Papaloizou}, J. 1986, \apj, 309, 846

\bibitem[{{Masset}(2001)}]{2001ApJ...558..453M}
{Masset}, F.~S. 2001, \apj, 558, 453

\bibitem[{{Masset}(2002)}]{2002A&A...387..605M}
{Masset}, F.~S. 2002, \aap, 387, 605

\bibitem[{{Masset} {et~al.}(2006{\natexlab{a}}){Masset}, {D'Angelo}, \&
  {Kley}}]{2006astro.ph..7155M}
{Masset}, F.~S., {D'Angelo}, G., \& {Kley}, W. 2006{\natexlab{a}}, ApJ accepted

\bibitem[{{Masset} {et~al.}(2006{\natexlab{b}}){Masset}, {Morbidelli}, {Crida},
  \& {Ferreira}}]{2006ApJ...642..478M}
{Masset}, F.~S., {Morbidelli}, A., {Crida}, A., \& {Ferreira}, J.
  2006{\natexlab{b}}, \apj, 642, 478

\bibitem[{{Masset} \& {Papaloizou}(2003)}]{2003ApJ...588..494M}
{Masset}, F.~S. \& {Papaloizou}, J.~C.~B. 2003, \apj, 588, 494

\bibitem[{{Menou} \& {Goodman}(2004)}]{2004ApJ...606..520M}
{Menou}, K. \& {Goodman}, J. 2004, \apj, 606, 520

\bibitem[{{Nelson} \& {Papaloizou}(2004)}]{2004MNRAS.350..849N}
{Nelson}, R.~P. \& {Papaloizou}, J.~C.~B. 2004, \mnras, 350, 849

\bibitem[{{Ogilvie} \& {Lubow}(2003)}]{2003ApJ...587..398O}
{Ogilvie}, G.~I. \& {Lubow}, S.~H. 2003, \apj, 587, 398

\bibitem[{{Paardekooper} \& {Mellema}(2006)}]{2006A&A...450.1203P}
{Paardekooper}, S.-J. \& {Mellema}, G. 2006, \aap, 450, 1203

\bibitem[{{Pollack} {et~al.}(1996){Pollack}, {Hubickyj}, {Bodenheimer},
  {Lissauer}, {Podolak}, \& {Greenzweig}}]{1996Icar..124...62P}
{Pollack}, J.~B., {Hubickyj}, O., {Bodenheimer}, P., {et~al.} 1996, Icarus,
  124, 62

\bibitem[{{Rafikov}(2002)}]{2002ApJ...572..566R}
{Rafikov}, R.~R. 2002, \apj, 572, 566

\bibitem[{{Tanaka} {et~al.}(2002){Tanaka}, {Takeuchi}, \&
  {Ward}}]{2002ApJ...565.1257T}
{Tanaka}, H., {Takeuchi}, T., \& {Ward}, W.~R. 2002, \apj, 565, 1257

\bibitem[{{Terquem}(2003)}]{2003MNRAS.341.1157T}
{Terquem}, C.~E.~J.~M.~L.~J. 2003, \mnras, 341, 1157

\bibitem[{{Ward}(1997)}]{1997Icar..126..261W}
{Ward}, W.~R. 1997, Icarus, 126, 261

\end{thebibliography}

\end{document}